\RequirePackage{ifpdf}
\ifpdf 
\documentclass[pdftex]{sigma}
\else
\documentclass{sigma}
\fi

\begin{document}
\allowdisplaybreaks

\renewcommand{\PaperNumber}{024}

\def\f{\varphi}
\def\e{\varepsilon}
\def\p{\partial}

\FirstPageHeading

\ShortArticleName{On the Essential Spectrum of Pseudorelativistic
Hamiltonians}

\ArticleName{On the Essential Spectrum\\
of Many-Particle Pseudorelativistic Hamiltonians\\
with Permutational Symmetry Account}

\Author{Grigorii ZHISLIN}

\AuthorNameForHeading{G.~Zhislin}

\Address{Radiophysical Research Institute,
 25/14   Bol'shaya Pechorskaya Str.,\\ Nizhny Novgorod, 603950 Russia}

\Email{\href{mailto:greg@nirfi.sci-nnov.rup}{greg@nirfi.sci-nnov.ru}}

\ArticleDates{Received October 27, 2005, in f\/inal form February
07, 2006; Published online February 20, 2006}

\Abstract{In this paper we formulate our results on the essential
spectrum of many-particle pseudorelativistic Hamiltonians without
magnetic and external potential f\/ields in the spaces of
functions, having arbitrary type $\alpha$ of the permutational
symmetry. We discover location of the essential spectrum for all
$\alpha$ and for some cases we establish  new properties of the
lower bound of this spectrum, which are useful for  study of the
discrete spectrum.}


\Classification{35P20; 35Q75; 46N50; 47N50; 70H05; 81Q10}

In this paper we formulate our results on the essential spectrum
of many-particle pseudore\-la\-tivistic Hamiltonians without
magnetic and external potential f\/ields in spaces of functions,
having arbitrary type $\alpha$ of the permutational symmetry. We
discover the location of the essential spectrum for all $\alpha$
(Theorem~1) and for some cases we establish new properties of the
lower bound of this spectrum, which are useful for study of the
discrete spectrum (Lemma~1).

Before this work similar results on the essential spectrum were
obtained in~\cite{damak,lewis}, but in~\cite{lewis} not arbitrary
$\alpha$ were considered, and the construction of the operator of
the relative motion was not invariant with respect to the
permutations of identical particles in contrast to our approach
(in this respect connection of our results with~\cite{lewis} is
the same, as  connection~\cite{sigalov} with~\cite{zhislin});
in~\cite{damak} more extensive class of pseudorelativistic
Hamiltonians was studied as compared to~\cite{lewis} and to this
paper, but in~\cite{damak} the permutational symmetry was not
considered. Moreover, our Lemma~1 is new.

{\bf 1.} Let $Z_1=\{0,1,\ldots,n\}$ be the quantum system of
$(n+1)$ particles, $m_i$, $r_i=(x_i,y_i,z_i)$ and $p_i$ be the
mass, the radius-vector and the momentum of $i$-th particle.
Pseudorelativistic~(PR) energy operator of $Z_1$ can be written in
the form
\begin{gather*}
{\cal H}'=K'(r)+V(r),
\end{gather*}
where $r=(r_0,r_1,\ldots,r_n)$,
\begin{gather*}
K'(r)=\sum_{j=0}^n\sqrt{-\Delta_j+m_j^2}^{\;\footnotemark},\qquad
V(r)=V_0(r)={1\over2} \sum_{i,j=0,\,i\ne j}^n\, V_{ij}(|r_{ij}|),
\end{gather*}
\footnotetext{We have chosen the unit system so the Plank constant
and the light velocity are equal to 1.} $\Delta_j={\p^2\over\p
x_j^2}+{\p^2\over\p y_j^2}+ {\p^2\over\p z_j^2}$,
$V_{ij}(|r_{ij}|)=V_{ji}(|r_{ji}|)$ be the real potential of the
interaction $i$-th and $j$-th particles, $r_{ij}=r_i-r_j$,
$V_{ij}(|r_1|)\in{\cal L}_{2,{\rm loc}}({\mathbb R}^3)$,
$V_{ij}(|r_1|)\to0$ at $|r_1|\to\infty$, and $V_{ij}(|r_{ij}|)$
are such that for some $\e_0 > 0$ operator  ${\cal H}'$ is
semibounded from below for $V(r)=(1+\e_0) V_0(r)$. If the system
$Z_1$ is a molecule, the last condition means that we may consider
only the molecules consisting of atoms of such elements whose
number in Mendeleev periodic table is smaller
than~85~\cite{lewis,lieb}.

The operator ${\cal H}'$ is not local: in the coordinate space
operators $\sqrt{-\Delta_j+m_j^2}$ are integral operators, in the
momentum representation multiplicators $V_{ij}(|r_{ij}|)$ turn
into integral operators. But in the momentum space the operators
$\sqrt{-\Delta_j+m_j^2}$ are multiplication operators. Actually,
let $p_j=(p_{j1},p_{j2},p_{j3})$, $p=(p_0,\ldots, p_n)$, $\f(r)\in
{\cal L}_2({\mathbb R}^{3n+3})$,                       and
$\overline\f(p)$ be Fourier-transform of~$\f(r)$:
\[
\overline\f(p)={1\over {(\sqrt{2\pi}\,)}^{3n+3}} \int_{{\mathbb
R}^{3n+3}}\,\f(r)\,e^{i(p,r)}\,dr,\] then
\[
\sqrt{-\Delta_j+m_j^2}\,\f(r)=\sqrt{p_j^2+m_j^2}\,\overline\f(p).\]
Let
\[
T'_j(p_j)=\sqrt{p_j^2+m_j^2},\qquad T'(p)=\sum_{j=0}^n T'_j(p_j).
\]
Now we can rewrite operators ${\cal H}'$ using mixed form writing:
\begin{gather*}
{\cal H}'=T'(p)+V(r),
\end{gather*}
where operators $T'(p)$ and $V(r)$ act in the momentum and in the
coordinate spaces respectively.

{\bf 2.} The operator ${\cal H}'$ corresponds to the energy of the
whole system motion. But for applications it is interesting to
know the spectrum of the operator corresponding to the {\em
relative\/} motion energy. To get such operator for
nonrelativistic~(NR) case one separate the center-of-mass motion,
but for pseudorelativistic~(PR) case it is impossible. To
construct the operator of the relative motion from PR operator
${\cal H}'$, we reduce the operator ${\cal H}'$ to any f\/ixed
eigenspace of operator of the total momentum~\cite{lewis}. Let
$\xi_0=(\xi_{01},\xi_{02},\xi_{03})$ be the center-of-mass
radius-vector:
\[
\xi_0=\sum_{j=0}^n m_jr_j/M,\qquad M=\sum_{j=0}^n m_j,
\]
$q_j=r_j-\xi_0$ be the relative coordinates of $j$-th particle,
$j=0,1,\ldots,n$, $q=(q_0,\ldots,q_n)$. We take $q$, $\xi_0$ as
the new coordinates of the particles from $Z_1$. Let us note that
vectors $q_0,\ldots,q_n$ are dependent: they belong to the space
\[
R_0=\left\{ q'\mid q'=(q'_0,\ldots,q_n'),\ \ \sum_{j=0}^n
m_jq_j'=\theta=(0,0,0)\right\}
\]
of relative motion. On the other hand, if
$q'=(q_0',\ldots,q_n')\in R_0$ and $\xi_0'$ is an arbitrary
f\/ixed vector from ${\mathbb R}^3$, we may consider $q_j'$ and
$\xi_0'$ as the relative coordinates of the point
$r_j'=q_j'+\xi_0'$, $j=0,1,\ldots,n$ and the center-of-mass
position of $Z_1$ respectively. It is easy to see that
Fourier-conjugate coordinates to $q_j$ are the same $p_j$ as for
$r_j$, and Fourier-conjugate coordinate for $\xi_0$ is ${\cal
P}_0=({\cal P}_{01},{\cal P}_{02},{\cal P}_{03})=
\sum\limits_{j=0}^n p_j$.

Let us consider the operators
\[
L_{0s}={1\over i}\,{d\over d\xi_{0s}},\qquad s=1,2,3.\] In the
momentum space these operators are multiplication operators
\[
\overline L_{0s}={\cal P}_{0s}.\] It follows from above that the
operators $L_{0s}\{\overline L_{0s}\}$ commute with ${\cal H}'$.
So any eigenspaces of the operators $L_{0s}$ are invariant for
${\cal H}'$. Let ${-}Q_{0s}$ be a real eigenvalue of the operator
$L_{0s}$, $W_{0s}$ be corresponding eigenspace and
\[
W_0=W_{01}\cap W_{02}\cap W_{03}.\] The space $W_0$ is invariant
for ${\cal H}'$. Evidently
\[
W_0=\left\{{(2\pi)}^{-3/2}\,e^{-i(Q_0,\xi_0)}\,\f(q)
\right\}^{\,\footnotemark},\qquad \overline
W_0=\left\{\overline\f(p)\, \prod\limits_{s=1}^3\delta({\cal
P}_{0s}-Q_{0s})\right\},\] where $Q_0=(Q_{01}, Q_{02}, Q_{03})$,
$\f(q)$ is an arbitrary function, $\f(q)\in{\cal L}_2(R_0)$, and
$\overline W_0$ is Fourier-image of $W_0$.

\footnotetext{The coef\/f\/icient ${(2\pi)}^{-3/2}$ in front of
the $e^{-i(Q_0,\xi_0)}$ plays the role of ``normalizing factor'':
Fourier-image of ${(2\pi)}^{-3/2}\,e^{-i(Q_0,\xi_0)}$ is
$\prod\limits_{s=1}^s\delta({\cal P}_{0s}-Q_{0s})$ without any
factor.}

Let us rewrite  operator ${\cal H}'$ using the coordinates $q$,
$\xi_0$ $\{p,{\cal P}\}$ and reduce it to the subspace
$W_0\{\overline W_0\}$. Then we obtain the operator ${\cal H}'$ in
the form
\begin{gather*}
H_0'=T'(p,Q_0)+V(q),
\end{gather*}
where
\[
T'(p,Q_0)=T'(p),\] but with the condition
\begin{gather}\label{z5}
\sum_{j=0}^n p_j=Q_0;
\\
V(q)={1\over2} \sum_{i,j=0,\,i\ne j}^n V_{ij}(|q_i-q_j|),\qquad
q_i-q_j=r_i-r_j.\nonumber
\end{gather}
We see that $H_0'$ depends on the relative coordinates $q$, their
momenta $p$ and the total momentum value $Q_0$. So if we f\/ix
$Q_0$ we obtain the operator, which can be considered as the
operator of the relative motion. We shall study this operator in
the space ${\cal L}_2(R_0)$ with condition~(\ref{z5}) for momenta.

For technical reasons it is convenient to take
\[
T_j(p_j)=T_j'(p_j)-m_j\] instead of $T_j'(p_j)$ and
\[
T(p,Q_0)=\sum_{j=0}^n T_j(p_j)\] instead of $T'(p;\,Q_0)$. So the
subject of our study is  operator
\begin{gather}\label{z6}
H_0=T(p;\,Q_0)+V(q)
\end{gather}
(with condition~(\ref{z5})). The operator $H_0$ is bounded from
below on $C_0^\infty(R_0)$. We extend it to a self-adjoint one
using Friedrichs extension, and save the notation $H_0$ for the
obtained operator.

Let us note that instead of the dependent coordinates
$q_0,\ldots,q_n$ we could introduce independent relative
coordinates (and their momenta) similar to~[2], but such approach
generates dif\/f\/iculties, when one takes into account the
permutational symmetry (see~\S~{\bf 5}), and we do not use this
approach.

{\bf 3.} We shall study spectrum of the operator $H_0$ not in the
whole space ${\cal L}_2(R_0)$, but in the subspaces of functions
from ${\cal L}_2(R_0)$, having the f\/ixed types of permutational
symmetry. We do this
\begin{enumerate}\itemsep=0pt
\item[i)] to satisfy the Pauli exclusion principle, \item[ii)] to
obtain additional information about the structure of the spectrum
$H_0$.
\end{enumerate}

We denote by $S$ and $\alpha$ the group of the permutations of all
identical particles of $Z_1$ and an arbitrary type of irreducible
representation of $S$ respectively. Let us determine the
operators~$T_g$, $g\in S$ by relations
\begin{gather*}
T_g\,\f(q)=\f(g^{-1}q),\qquad g\in S
\end{gather*}
and put
\[
P^{(\alpha)}={l_\alpha\over |S|}\sum\limits_{g\in S} \overline
\chi_g^{(\alpha)}\,T_g,\qquad B^{(\alpha)}=P^{(\alpha)}\,{\cal
L}_2(R_0),\] where $\chi_g^{(\alpha)}$ is the character of the
element $g\in S$ in the irreducible representation of the
type~$\alpha$, $l_\alpha$~is the dimension of this representation,
$|S|$ is the number of elements of $S$. The operator
$P^{(\alpha)}$ is the projector in ${\cal L}_2(R_0)$ on the
subspace $B^{(\alpha)}=B^{(\alpha)}(R_0)$ of functions, which are
transformed by the operators $T_g$, $g\in S$, according to the
representation of the type $\alpha$~\cite{wigner}. Evidently
$P^{(\alpha)}H_0= H_0P^{(\alpha)}$. Let
$H_0^{(\alpha)}=H_0P^{(\alpha)}$. $H_0^{(\alpha)}$ be the
restriction of the operator $H_0$ to the subspace $B^{(\alpha)}$
of functions, having the permutational symmetry of the type
$\alpha$.

In this paper we discover location of the essential spectrum
$s_{\rm ess}\big(H_0^{(\alpha)}\big)$ of the operator
$H_0^{(\alpha)}$.

{\bf 4.} Let $Z_2=(D_1,D_2)$ be an arbitrary decomposition of the
initial system $Z_1$ into 2 non-empty clusters $D_1$ and $D_2$
without common elements:
\[
D_1\cup D_2=Z_1,\qquad D_1\cap D_2=\varnothing\] and
\begin{gather}\label{z8}
H(Z_2)=T(p,Q_0)+V(q;Z_2),
\end{gather}
where
\[
V(q;Z_2)={1\over2}\sum_{s=1}^2\, \sum_{i,j\in D_s,\,i\ne
j}V_{ij}(|q_j-q_i|).\] $H(Z_2)$ is the PR energy operator of
compound system $Z_2$, consisting of non interacting (one with
other) clusters $D_1$, $D_2$ with the same condition~(\ref{z5})
for the total momentum as for~$Z_1$:
\[
\sum_{i=0}^n p_i=Q_0.\] Let $S[D_s]$ be the group of the
permutations of all identical particles from $D_s$, $s=1,2$, $\hat
g$ be the permutation $D_1\leftrightarrow D_2$ if these clusters
are identical $(D_1\sim D_2)$. We put
\begin{gather*}
S_0(Z_2)=S[D_1]\times S[D_2],\\
S(Z_2)=S_0(Z_2) \quad \hbox{if}\quad D_1\not\sim D_2,\\
S(Z_2)=\hat S(Z_2)=S_0(Z_2)\cup S_0(Z_2)\hat g\quad \hbox{if}\quad
D_1\sim D_2.
\end{gather*}
$S(Z_2)$ is the group of the permutational symmetry of the
compound system $Z_2$. It is clear that $S_0(Z_2)\subseteq
S(Z_2)\subseteq S$.

Let $ F(\alpha;Z_2)\!=\!\{\alpha'\}
\big\{F_0(\alpha;Z_2)\!=\!\{\check\alpha\} \big\}$ be the set of
all types $\alpha'\{\check\alpha\}$ of the group
$S(Z_2)\{S_0(Z_2)\}$ irreducible representations, which are
contained in the group $S$ irreducible representation
$D_g^{(\alpha)}$ of the type $\alpha$ after reducing
$D_g^{(\alpha)}$ from $S$ to $S(Z_2)\{S_0(Z_2)\}$. For
$\forall\;\alpha'\{\check\alpha\}$ we determine the projector
$P^{(\alpha')}(Z_2)\{P^{(\check\alpha)}(Z_2)\}$ on the subspace of
functions $\f(q)$, which are transformed by  operators $T_g$
\[
T_g\,\f(q)=\f(g^{-1}q),\qquad g\in S(Z_2),\qquad \{g\in
S_0(Z_2)\}\] according to the group $S(Z_2)\{S_0(Z_2)\}$
irreducible representation of the type $\alpha'\{\check\alpha\}$.

Let $\gamma=\alpha'$ or $\gamma=\check\alpha$; obviously if
$P^{(\gamma)}(Z_2)\,\f(q)=\f(q)$, then
$P^{(\gamma)}(Z_2)\,\overline\f(p)=\overline\f(p)$. We set
\begin{gather*}
P(\alpha;Z_2)=\sum_{\alpha'\in F(\alpha;Z_2)}
P^{(\alpha')}(Z_2),\qquad \check
P(\alpha;Z_2)=\sum_{\check\alpha\in F_0(\alpha;Z_2)}
P^{(\check\alpha)}(Z_2),\\
H(\alpha;\,Z_2)=H(Z_2)P(\alpha;Z_2),\qquad \check
H(\alpha;Z_2)=H(Z_2)\,\check P(\alpha;Z_2).
\end{gather*}
The operator $H(\alpha;Z_2)\{\check H(\alpha;Z_2)\}$ is the
restriction of the operator $H(Z_2)$ (see~(\ref{z8})) to the
subspace $B(\alpha;Z_2)=P(\alpha;Z_2)\,{\cal L}_2(R_0)$ $\{\check
B(\alpha;Z_2)= \check P(\alpha;Z_2)\,{\cal L}_2(R_0)\}$. Let
\begin{gather*}
\mu^{(\alpha)}=\min\limits_{Z_2}\,\inf H(\alpha;Z_2).
\end{gather*}
It is possible to prove that
\begin{gather}\label{z10}
\mu^{(\alpha)}=\min\limits_{Z_2}\,\inf \check H(\alpha;Z_2).
\end{gather}
We denote by $A(\alpha)$ the set of all $Z_2$, for which
\[
\inf\check H(\alpha;Z_2)=\min\limits_{Z_2'}\inf\check
H(\alpha;Z_2');\] then
\begin{gather}\label{z11}
\mu^{(\alpha)}=\inf\check H(\alpha;Z_2),\qquad Z_2\in A(\alpha).
\end{gather}

{\bf 5.} Our main result is the following theorem

\begin{theorem}
Essential spectrum $s_{\rm ess}\big(H_0^{(\alpha)}\big)$ of the
operator $H_0^{(\alpha)}$ consists of all points half-line
$[\mu^{(\alpha)},+\infty)$.
\end{theorem}

Let us compare  Theorem 1 with the corresponding results
in~\cite{lewis}.

First, in~\cite{lewis} a similar result was proved only for one of
simplest types $\alpha$ of the permutational symmetry (for
$\alpha$ corresponding to one-column Young scheme), while here we
assume arbitrary~$\alpha$.

Second, we use more natural, simple and transparent approach for
taking symmetry into account, compared to~\cite{lewis}. Actually,
we apply relative coordinates~$q_i$ with respect to center-of-mass
position $\xi_0$: $q_i=r_i-\xi_0$, $i=0,1,\ldots,n$ and so the
transposition $g_j$: $r_j\leftrightarrow r_0$ of $j$-th and $0$-th
particles results in the transposition of $q_j$ and $q_0$ only,
but just as all other coordinates~$q_i$, $i\ne j$, $i\ne 0$, are
without any change. In~\cite{lewis}  relative coordinates~$\tilde
q_i$ are taken with respect to the position of $0$-th particle:
$\tilde q_i=r_i-r_0$, $i=1,2,\ldots,n$ and this choice implies
changing of all $\tilde q_i$ under transposition $g_j$. Namely,
$T_{g_j}\,\psi(\tilde q)=\psi(g_j^{-1}\,\tilde q)=\psi(\hat q)$,
where $\tilde q=(\tilde q_1,\ldots,\tilde q_n)$, $\hat q=(\hat
q_1,\ldots\hat q_n)$, $\hat q_i=\tilde q_i-\tilde q_j$, $i\ne j$,
$\hat q_j=-\tilde q_j$. Such situation is not realized only if the
system~$Z_1$ contains a particle, which is not identical to any
other particle from $Z_1$ (and if we index this particle by
number~0), but there is no such exceptional particle in the most
number of molecules. Completing the second remark, we can note,
roughly speaking, that our approach for taking permutational
symmetry into account follows~\cite{sigalov}, while
authors~~\cite{lewis} follow~\cite{zhislin}.

{\bf 6.} We do not write here the proof of the Theorem~1, since
the signif\/icant part of this proof will be needed for the study
of the discrete spectrum $s_d\big(H_0^{(\alpha)}\big)$ of the operator
$H_0^{(\alpha)}$ (this study is not f\/inished), so we shall
publish the full proof of the Theorem~1 later (together with the
results on the discrete spectrum). But here we shall do some
preparations for our next paper. Namely, we shall obtain
from~(\ref{z10}), (\ref{z11}) the other formula for
$\mu^{(\alpha)}$, which is more convenient for the investigation
of the structure $s_d\big(H_0^{(\alpha)}\big)$. To do it f\/irst
of all we transform the expression of the operator $H(Z_2)$ for
f\/ixed $Z_2=(D_1,D_2)$. We introduce  clusters $D_s$
center-of-mass coordinates
\[
\xi_s=(\xi_{s1},\xi_{s2},\xi_{s3})=\sum_{j\in
D_s}\,r_jm_j/M_s,\qquad M_s=\sum_{j\in D_s} m_j,\] the relative
coordinates $q_j(Z_2)=r_j-\xi_s$, $j\in D_s$, of the particles
from $D_s$ with respect to center-of-mass position and the vector
$\eta=\xi_2-\xi_1$. Evidently, $q_j(Z_2)=q_j+\xi_0-\xi_s$, where
$\xi_0-\xi_1=M_2\eta/M$, $\xi_0-\xi_2=-M_1\eta/M$.

The coordinates $q(Z_2)=\big(q_0(Z_2),\ldots,q_n(Z_2)\big)$ are
not independent, since $\sum\limits_{j\in
D_s}m_j\,q_j(Z_2)=\theta$, $s=1,2$. It is easy to see that
Fourier-conjugate coordinates to $q_j(Z_2)$ are the same $p_j$
that were introduced before. Let ${\cal P}_s=\sum\limits_{j\in
D_s}p_j$. Then Fourier-conjugate coordinates to $\eta$ are
\begin{subequations}\label{z12}
\begin{gather}\label{z12a}%
{\cal P}_\eta=({\cal P}_{\eta1},{\cal P}_{\eta2},{\cal
P}_{\eta3})= ({\cal P}_2M_1-{\cal P}_1M_2)/M
\end{gather}
where by~(\ref{z5})
\begin{gather}\label{z12b}
{\cal P}_1+{\cal P}_2=Q_0.
\end{gather}
\end{subequations}
We consider $q(Z_2)$ and $\eta$ as new coordinates of particles
from $Z_1$ and denote the operator $H(Z_2)$ in new coordinates by
$H_0(Z_2)$. According to consideration above and since
$q_i-q_j=q_i(Z_2)-q_j(Z_2)$, $i,j\in D_s$, $s=1,2$, we have
\begin{gather}\label{z13}
H(Z_2)=H_0(Z_2)=T(p,Q_0,{\cal P}_\eta)+V\big(q(Z_2);Z_2\big)
\end{gather}
where the operator~(\ref{z13}) has the same form as the
operator~(\ref{z8}), but the conditions~\eqref{z12} have to be
satisfied.

Let us introduce  spaces
\begin{gather*}
R_0(Z_2)=\left\{ q(Z_2)\mid
q(Z_2)=\big(q_0(Z_2),\ldots,q_n(Z_2)\big),\quad
\sum_{j\in D_s} m_j\,q_j(Z_2)=\theta,\quad s=1,2\right\},\\
R_\eta=\left\{\eta\mid\eta=(\eta_1,\eta_2,\eta_3)\right\},\qquad
R_{0,\eta}(Z_2)=R_0(Z_2)\,\oplus\,R_\eta,\\
{\cal L}_2\big(R_{0,\eta}(Z_2)\big)=\left\{
\f\big(q(Z_2),\eta\big)\; \Big | \; \int_{R_{0,\eta}}
{|\f|}^2dq(Z_2)\,d\eta<+\infty\right\}.
\end{gather*}
In the space ${\cal L}_2\big(R_{0,\eta}(Z_2)\big)$ we determine
operators $P_0^{(\check\alpha)}(Z_2)$ similarly to operators
$P^{(\check\alpha)}(Z_2)$, but now the operators $T_g$, $g\in
S_0(Z_2)$, are def\/ined on functions $\f\big(q(Z_2),\eta\big)$
and $\overline\f(p,{\cal P}_\eta)$ by  relations
\[
T_g\,\f\big(q(Z_2),\eta\big):=
\f\big(g^{-1}\,q(Z_2),\eta\big),\qquad T_g\,\overline\f(p,{\cal
P}_\eta)= \overline\f(g^{-1}p,{\cal P}_\eta).\] Here we took into
account that $g^{-1}\eta=\eta$ and $g^{-1}{\cal P}_\eta={\cal
P}_\eta$ for $\forall\;\eta,{\cal P}_\eta$, $\forall \; g\in
S_0(Z_2)$.

Let us
\[
\check P_0(\alpha;Z_2)= \sum_{\check\alpha\in
F_0(\alpha;Z_2)}\,P_0^{(\check \alpha)}(Z_2), \qquad \check
H_0(\alpha;\,Z_2)= H_0(Z_2)\,\check P_0(\alpha;Z_2).\] According
to~(\ref{z11}),
\[
\mu^{(\alpha)}=\inf\check H_0(\alpha;Z_2),\qquad Z_2\in
A(\alpha),
\] where the operator $\check H_0(\alpha;Z_2)$ is
considered in the space ${\cal L}_2(R_{0,\eta})$. Since the
operator $T(p,Q_0,{\cal P}_\eta)$ is a multiplication operator and
the potential $V\big(q(Z_2);Z_2\big)$ does not depend on $\eta$,
we may consider the operator $\check H_0(\alpha;Z_2)\equiv\check
H_0(\alpha;Z_2;{\cal P}_\eta)$ in the space ${\cal
L}_2\big(R_0(Z_2)\big)$ at the arbitrary f\/ixed ${\cal
P}_\eta=Q$. Then
\begin{gather}\label{z14}
\mu^{(\alpha)}=\inf\limits_Q\inf\check H_0(\alpha;Z_2;Q),\qquad
Z_2\in A(\alpha).
\end{gather}
Operator $\check H(\alpha;Z_2;Q)$ depends on $Q$ continuously and
\[
\lim\limits_{|Q|\to+\infty}\inf\check H_0(\alpha;Z_2;Q)=
+\infty,\] since if $|Q|\to+\infty$, then at least for one $j$ it
holds $|p_j|\to\infty$ and consequently $T(p,Q_0,Q)\to+\infty$. So
there exists a compact set $\Gamma(\alpha;Z_2)$ of such vectors
$Q\in {\mathbb R}^3$ that
\begin{gather*}
\mu^{(\alpha)}=\inf\check H_0(\alpha;Z_2;Q),\qquad
Q\in\Gamma(\alpha;Z_2),\qquad Z_2\in A(\alpha).
\end{gather*}

{\bf 7.} Unfortunately, in the general case we know nothing about
f\/initeness or inf\/initeness of the number of the set
$\Gamma(\alpha;Z_2)$ elements. But we can prove the following
assertion

\begin{lemma}
Let for some open region $W\subset {\mathbb R}^3$,
$\Gamma(\alpha;Z_2)\subset W$,

\begin{enumerate}\itemsep=0pt
\item[\rm i)] $\lambda(\alpha;Z_2;Q):=\inf\check
H_0(\alpha;Z_2;Q)$ is the point of the discrete spectrum of the
operator \linebreak $\check H_0(\alpha; Z_2;Q)$ for $Q\in W$,
\item[\rm ii)] there is such $\check\alpha_0$, which does not
depend on $Q$, that the representation $g\to T_g$, $g\in S_0(Z_2)$
in the eigenspace $U(\alpha;Z_2;Q)$ of the operator $\check
H_0(\alpha;Z_2;Q)$, corresponding to its eigenvalue
$\lambda(\alpha;Z_2;Q)$, has ONE irreducible component of the type
$\check\alpha_0$ for each $Q\in W$.
\end{enumerate}
Then the set $\Gamma(\alpha;Z_2)$ is finite.
\end{lemma}

\begin{proof}
Let $\check B_0(\alpha;Z_2)=\check P_0(\alpha;Z_2) {\cal
L}_2\big(R_0(Z_2)\big)$. Since
\[
\check P_0(\alpha;Z_2)=P_0^{(\check\alpha_0)}(Z_2)+
\sum\limits_{\check\alpha\in
F_0(\alpha;Z_2),\check\alpha\ne\check\alpha_0}\,
P_0^{(\check\alpha)}(Z_2),\] then
\[
B_0^{(\check\alpha_0)}(Z_2):= P_0^{(\check\alpha_0)}(Z_2)\,\check
B_0(\alpha;Z_2)= P_0^{(\check\alpha_0)}(Z_2)\,{\cal
L}_2\big(R_0(Z_2)\big).\] It follows from the conditions i), ii)
that in the space
\[
U^{(\check\alpha_0)}= U(\alpha;Z_2;Q)\cap
B_0^{(\check\alpha_0)}(Z_2)\equiv
P_0^{(\check\alpha_0)}\,U(\alpha;Z_2;Q)\] the representation $g\to
T_g$, $g\in S_0(Z_2)$ is irreducible and has the type
$\check\alpha_0$.

Let $P_{01}^{(\check\alpha_0)}$ be the projector in
$B_0^{(\check\alpha_0)}(Z_2)$ on the space
$B_{01}^{(\check\alpha_0)}(Z_2)$ of functions, which belong to the
f\/irst line of the group $S_0(Z_2)$ irreducible representation of
the type $\check\alpha_0$.

Then the space $B_{01}^{(\check\alpha_0)}(Z_2)$ is invariant under
the operator $H_0(Z_2)$ and in this space the minimal eigenvalue
 $\lambda(\alpha;Z_2;Q)$ of the operator $H_0(Z_2)$ is
nondegenerated, since the corresponding eigenspace
$P_{01}^{(\check\alpha_0)}\,U^{(\check\alpha_0)}$ is
one-dimensional. In other words, the minimal eigenvalue of the
opera\-tor $P_{01}^{(\check\alpha_0)}\,H_0(\alpha;Z_2;Q)$ is
nondegenerated at $\forall\, Q\in W$. But if
$\lambda(\alpha;Z_2;Q)$ is nondegenerated, then
$\lambda(\alpha;Z_2;Q)$ is analytical function of $Q$, since the
operator $H_0(Z_2)$ is analytical function on~$Q$~\cite{reed}.
That is why there is only  f\/inite number of such vectors $Q$,
for which
\begin{gather*}
\mu^{(\alpha)}=\lambda(\alpha;Z_2;Q).\tag*{\qed}
\end{gather*}\renewcommand{\qed}{}
\end{proof}

{\bf 8. Discussion.} Theorem~1 and Lemma~1 describe the location
of essential spectrum $s_{\rm ess}\big(H_0^{(\alpha)}\big)$ of the
operator $H_0^{(\alpha)}$  and some properties of its lower bound
respectively. Now let us consider a role of these results for the
discrete spectrum study. It follows from Theorem~1 that to prove
the existence of nonempty discrete  spectrum
$s_d\big(H_0^{(\alpha)}\big)$ of the operator $H_0^{(\alpha)}$ it
is  suf\/f\/icient  to construct such trial function $\psi$,
$P^{(\alpha)}\psi=\psi$ that
\begin{gather}\label{z17}
\left(H_0^{(\alpha)}\psi,\psi\right)<\mu^{(\alpha)}(\psi,\psi),
\end{gather}
where the number $\mu^{(\alpha)}$ is determined by the relations
\eqref{z11} and \eqref{z14}.
 Construction of a function  $\psi$ for \eqref{z17} is important component of
geometrical methods application in the study of the spectrum
$s_d\big(H_0^{(\alpha)}\big)$ of operator $H_0^{(\alpha)}$.

But Theorem 1 is not a suf\/f\/icient base to study the spectral
asymptotics of the discrete spectrum $s_d\big(H_0^{(\alpha)}\big)$
near $\mu^{(\alpha)}$, when this spectrum is inf\/inite. To
understand the reason for that, let us consider the case when
$\mu^{(\alpha)}$ is the point of the spectrum
$s_d\big(H(\alpha;Z_2;Q)\big)$ for $Z_2\in A(\alpha)$,
$Q\in\Gamma(\alpha;\,Z_2)$ (such situation is expected
 for PR atoms). Then the inf\/inite series of the eigenvalues
$\lambda_k(Q)$, $k=1,2,\ldots$, from $s_d\big(H_0^{(\alpha)}\big)$
 may exist for $\forall\;Q\in\Gamma(\alpha;Z_2)$.
In this case it is possible to show that corresponding
eigenfunctions
 $\psi_k$ describe (when $k\to\infty$) such decomposition
$Z_2=\{C_1,C_2\}$ of the initial system $Z_1$, for which
\[
{\cal P}_1+{\cal P}_2=Q_0,\qquad M_1{\cal P}_2-M_2{\cal P}_1=MQ
\]
(see \eqref{z12}). Consequently, if the set $\Gamma(\alpha;Z_2)$
is inf\/inite, then the spectrum $s_d\big(H_0^{(\alpha)}\big)$
may consist of inf\/inite number of the inf\/inite series
eigenvalues $\lambda_k(Q)$, $k=1,2,\ldots$, where all series are
determined by the value $Q$ from $\Gamma(\alpha;Z_2)$. For such
situation there are no approaches to get the spectral asymptotics
of $s_d\big(H_0^{(\alpha)}\big)$. Thus, it was very desirable to
establish the conditions of impossibility of this situation that
is the conditions of f\/initeness of the set $\Gamma(\alpha;Z_2)$.
Namely, such conditions are given in Lemma~1 of the paper.

\subsection*{Acknowledgements}
This investigation is supported by RFBR grant 05-01-00299.

\LastPageEnding

\end{document}